\documentclass[aps,twocolumn,prb,floatfix,showpacs,superscriptaddress]{revtex4}

\usepackage{amsmath}
\usepackage{amssymb}
\usepackage{amsfonts}
\usepackage[pdftex]{graphicx}
\usepackage{units}
\usepackage{bm}
\usepackage{color}
\usepackage{color}

\renewcommand{\vec}[1]{\ensuremath{\boldsymbol{#1}}}

\begin{document}
\title{Pinning of thermal excitations at defects in artificial spin-ice dipolar arrays:\\ A theoretical investigation}

\author{Danny Thonig}
\email{dthonig@mpi-halle.mpg.de}
\affiliation{Max-Planck-Institut f\"ur Mikrostrukturphysik, D-06120 Halle (Saale), Germany}
\affiliation{Institut f\"ur Physik, Martin-Luther-Universit\"at Halle-Wittenberg, D-06099 Halle (Saale), Germany}

\author{J\"urgen Henk}
\affiliation{Institut f\"ur Physik, Martin-Luther-Universit\"at Halle-Wittenberg, D-06099 Halle (Saale), Germany}

\date{\today}

\begin{abstract}
 In this theoretical investigation we address the effect of defects on thermal excitations in square-lattice dipolar arrays. The geometry of the nanomagnets, adopted from recent experiments [A. Farhan \textit{et al.}, Nature Phys.\ \textbf{9} (2013) 375], allows for thermal activation at room temperature. It turns out that excitations can efficiently be pinned at defects.  Furthermore, it is possible to produce ferromagnetic strings of nanomagnets that connect a pair of defects; their lengths are closely related to the distance of the defects and the topology of the magnetic ground-state configuration. We discuss various types of defects, all of which may be produced by microstructuring techniques.
\end{abstract}


\maketitle

\section{Introduction}
\label{sec:introduction}
Topologically frustrated systems offer unexpected properties which have been studied recently in magnetic materials. A particular exciting system is artificial spin ice with its `exotic' magnetic ground state \cite{Schiffer02,Hodges11,Hamann12,Nisoli13}. Artificial spin ice is a two-dimensional (2D) array of magnetic nanoislands fabricated with desired geometries. The nanoislands are typically elongated to show a single-domain state; their magnetic moments then point in one of two directions. Being isolated from each other, e.\,g., separated by a distance in the order of several hundred nanometer, they are coupled by the long-range dipole-dipole interaction.

The systems sketched above are known for low-temperature fractionalization: they exhibit collective excitations that carry only a fraction of the elementary constituent's properties. These appear due to the absence of a unique ground state and due to the violation of the `two-in-two-out' ice rule, the latter proposed by Pauling for the proton ordering in water ice \cite{Pauling35}. Collective excitations appear as ferromagnetically aligned nanoislands (`strings') with end points (`nodes' or `vortices') that behave like magnetic charges, known as Nambu magnetic monopoles \cite{Nambu77,Mol09,Mol10,Morgan10,Moeller09}. The oppositely charged end nodes of the strings interact with each other with a distance dependence of a Coulomb  potential \cite{Mol10}. The properties of these artificial magnetic monopoles are studied theoretically as well as experimentally with great effort. We recall that magnetic monopoles would `symmetrize' electrodynamics with respect to the electric and the magnetic field \cite{Mohammadi07,Song96}.

Improvements in nanolithography allow to design artificial spin ice in 2D structures like honeycomb (kagom\'{e} ice) \cite{Tanaka06,Qi08,Moeller09, Nisoli}, brickwork \cite{Li10,Morrison13}, triangular \cite{Mol12}, and pentagonal \cite{Chern13} lattices. A three-dimensional artificial spin ice can be realized, e.\,g., by a layer-by-layer synthesis \cite{Chern13}. Local modifications in the otherwise perfect array are introduced by local nanolithography \cite{Garcia06,Buyukkose09}. These perturbations, considered as defects, modify significantly the properties of the system \cite{Esquinazi13,Andriotis13}. This leads to a question on the role of defects in dipolar arrays concerning thermal excitations. For three-dimensional spin ice and pyrochlore lattices, defects have already been studied by Jaubert \cite{Jaubert10}. First investigations of defects in 2D artificial spin ice have been performed by Silva et al. \cite{Silva13} who investigated the interaction between magnetic string excitations caused by defects and Nambu string excitations. Since defects can serve as pinning centers in magnetic materials\cite{Vansteenkiste08,Jourdan07}, it is conceivable to influence and control magnetic monopoles and the associated string excitations by tailoring the defects' properties.

Experimentally, effective thermal excitations in artificial square dipolar arrays can for example be provided by field protocols\cite{Nisoli07,Ke08} of vibro-fluidized granular matter, allowing  investigations of the short-range magnetic order. Long-range ordered ground states in square-lattice spin ice were obtained by thermal annealing during the fabrication \cite{Morgan10}, which is recommended for magnetic monopoles of charges $\pm 2$. Charges of $\pm 4$ were not observed, which is attributed to the nanoislands' geometry. Recent investigations on artificial spin ice with reduced nanoislands' dimension \cite{Farhan13} proved thermal excitations at room temperature, which allows not only to observe but also to control excitations with large magnetic charges.

In this work, we study various defect types in square-lattice artificial spin ice, with a focus on magnetic excitations. We consider modifications of two nanoislands of a vortex as a defect: \textit{(i)} removal of two islands from the vortex, \textit{(ii)} a vertical displacement of two islands, \textit{(iii)} a modification of the islands' thickness or \textit{(iv)} of their magnetization density. Furthermore, two defects with predefined positions and properties have been introduced into the spin ice. The resulting string excitations that link these defects are analyzed with respect to defect position and string length. Magnetic ground states have been achieved by Monte Carlo simulations for given temperatures \cite{Binder97b,Boettcher12}.

The paper is organized as follows. Theoretical aspects are presented in Section~\ref{sec:theory}. In the discussion of the results, given in Section~\ref{sec:discussion}, we address randomly distributed vacancies (\ref{sec:vacancies}) before analyzing one or two defects with prescribed properties (\ref{sec:defects}). We conclude with Section~\ref{sec:conclusion}.

\section{Theoretical aspects}
\label{sec:theory}
We build up the artificial spin ice by nanomagnets whose dimension has been taken from Ref.~\onlinecite{Farhan13} (length $\unit[470]{nm}$, width $\unit[170]{nm}$, and height $\unit[3]{nm}$). The lattice constant $a$ of the square lattice \cite{Wang06} is $\unit[793.8]{nm}$ (the lattice spacing in Ref.~\onlinecite{Farhan13a} is $\unit[425]{nm}$). The magnetic single-domain state of each elongated nanoisland is described by a magnetization vector $\pm \vec{M}$ (`spin') aligned along the large island axis. For islands made of permalloy, $|\vec{M}|\approx \unit[200 \cdot 10^{3}]{A m^{-1}}$.

\begin{figure}
 \centering
 \includegraphics[scale=0.48]{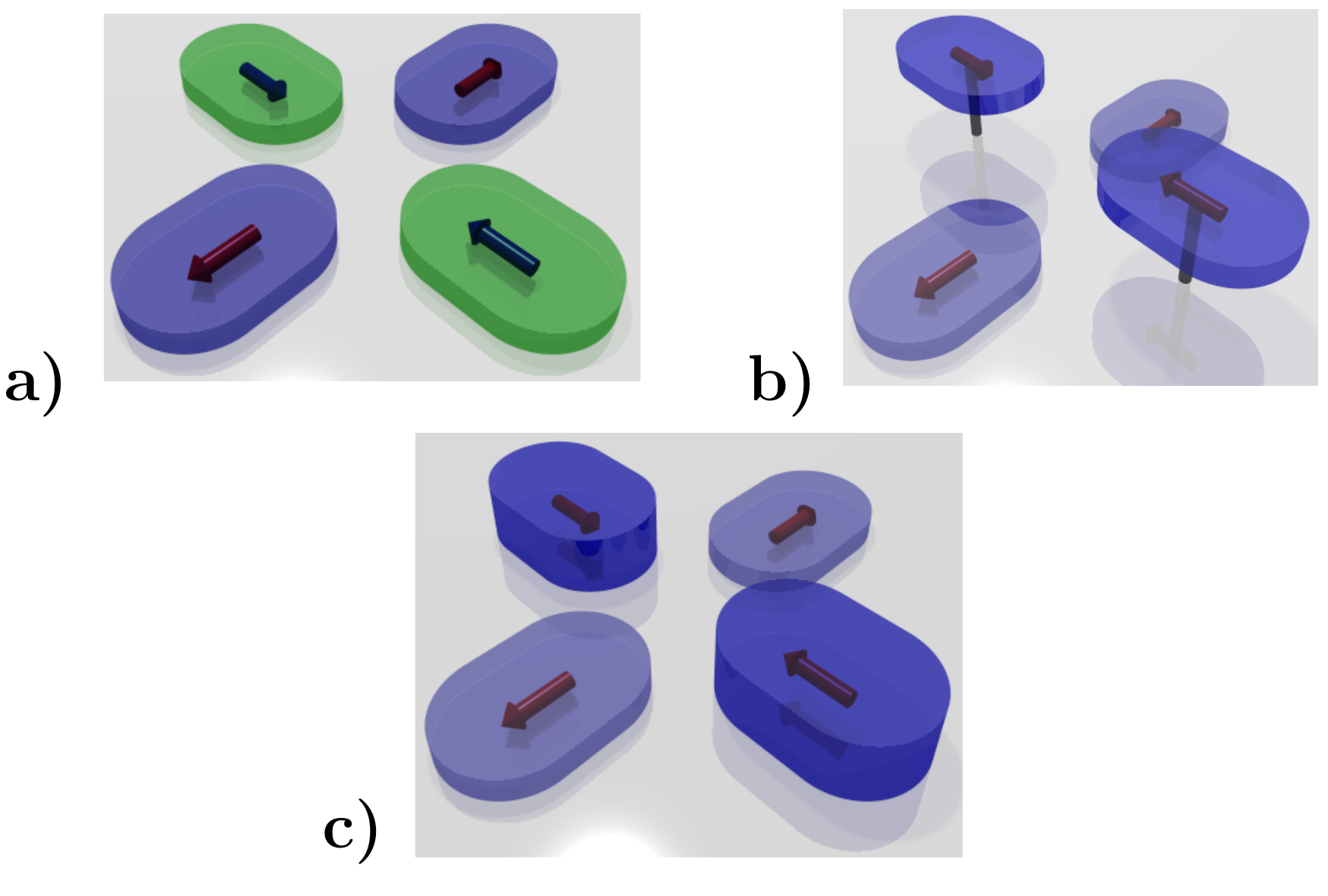}
 \caption{(Color online) Defects in a square-lattice dipolar array. At a selected node, opposite islands are modified by (a) a variation of their magnetization density, (b) a vertical displacement, and (c) a variation of their thickness. The modified islands are distinguished by color: green in case of (a), dark blue for (b) and (c). The arrows in each island indicate their magnetization $\vec{M}$.}
 \label{fig:defect}
\end{figure}

\begin{figure}
 \centering
 \includegraphics[width = 0.9\columnwidth]{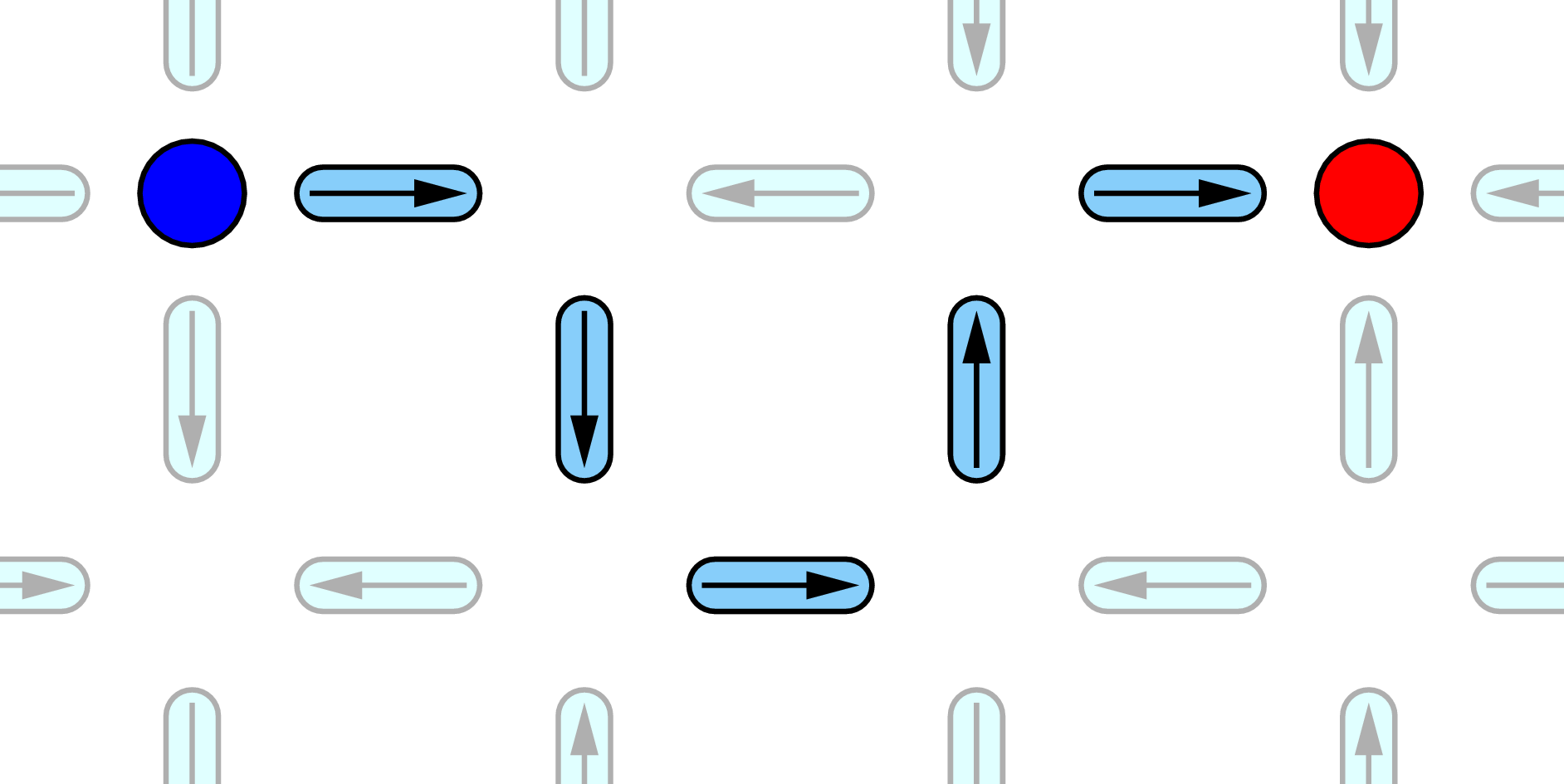}
 \caption{(Color online) String in a square-lattice dipolar array. The end points, carrying magnetic charges of $-4$ (dark blue dot, left-hand side) and $+4$ (red dot, right-hand side), are linked by a ferromagnetic path of islands (blue). The spin ice `host' is visualized by brighter colors. The arrows in each island indicate their magnetization $\vec{M}$.}
 \label{fig:string}
\end{figure}

Four islands that form a cross introduce a node (`vortex'; cf.\ Fig.~\ref{fig:defect}a). The number of spins pointing toward the node's center $\vec{C}$  define the charge $Q$ of that node ($Q \in \{ -4, -2, 0 , +2 , +4\}$; for example, $Q = 0$ in Fig.~\ref{fig:defect}a).  To quantify thermal activation, we introduce the fraction of nodes with charge $Q$  in the sample, $\eta_{Q} \equiv N_{Q} / N$ ($N$ number of nodes with four adjacent islands in the sample); on average $\langle \eta_{Q} \rangle = \langle \eta_{-Q} \rangle$. A path of ferromagnetically aligned nanoislands that connects a pair of nodes with opposite non-zero charges is called a `string excitation' (Fig.~\ref{fig:string}).

Instead approximating the nanomagnets as points\cite{Moeller06,Mol09,Mol10} or dipolar needles\cite{Moeller06,Moeller09}, we compute the dipole-dipole interaction energies for realistic shapes. This is done numerically, allowing in principle for arbitrarily shaped nanoislands. It turns out that the dipolar interaction\cite{Mengotti08,Mengotti09} is relevant only for first- (1NN) and for second-nearest (2NN) neighbors,\cite{Vedmedenko05,ThonigArXiv13} with energies $E_{\mathrm{1NN}}$ and $E_{\mathrm{2NN}}$, respectively.

The interaction energies $E_{\mathrm{1NN}}$ and $E_{\mathrm{2NN}}$ can be modified in various ways: by \textit{(i)} introducing vacancies (i.\,e., removal of two islands from a node), \textit{(ii)} modifying an island's magnetization density (Fig.~\ref{fig:defect}a), \textit{(iii)} varying the vertical displacement $\delta z$ (Fig.~\ref{fig:defect}b), and \textit{(iv)} increasing an island's thickness (Fig.~\ref{fig:defect}c). These modifications keep the charges $Q$ even. Local modifications of the free-energy landscape are achieved by introducing defects into the dipolar arrays.

To obtain a magnetic ground state, we use Monte Carlo simulations based on the Metropolis algorithm  \cite{Binder97b,Boettcher12}. The energy barrier $\Delta E$ between the two possible magnetization states $\pm \vec{M}$ of the nanoislands is considered. A slight reduction of the temperature in each step drives the system toward a free-energy minimum by successively reversing the islands' spins. A typical Monte Carlo simulation comprises at least 100\,000 steps. In this Paper, we report on results for lattices with $20 \times 20$ cells and with $50 \times 50$ cells, each with $2$ nanomagnets ($N_{\mathrm{spin}} = 20 \times 20 \times 2 = 800$ or $5000$). These samples are large enough to suppress even minute finite-size effects, as has been checked by comparison with calculations for larger arrays. 

The magnetic ground state of artificial spin ice at small finite temperature ($T\approx\unit[1]{K}$) is dictated by the (spin) ice rule: at a node with center $\vec{C}$, two spins pointing inward and two spins pointing outward (`2In2Out' rule, $Q = 0$). Hence, $\eta_{0} = \unit[100]{\%}$. Determined by the interaction energies $E_{\mathrm{1NN}}$ and $E_{\mathrm{2NN}}$, the ground state is `2in2outOp' for $\delta_{z} = 0$, in agreement with earlier work (e.\,g., Ref.~\onlinecite{Mol10}); it shows inward pointing spins at opposite (`Op') nanomagnets. This state shows a degree of degeneracy of $2$, with the energy $E = \left(-4 E_{\mathrm{1NN}} + 2 E_{\mathrm{2NN}}\right)  N_{\mathrm{spin}}$ ($N_{\mathrm{spin}}$ number of spins in the sample). The configuration `2In2OutAd' has an energy of $-2 E_{\mathrm{2NN}} N_{\mathrm{spin}}$,  is four-fold degenerate, and consists of  inward pointing spins at adjacent (`Ad') islands \cite{ThonigArXiv13}.

\begin{figure}
 \centering
 \includegraphics[scale=0.58]{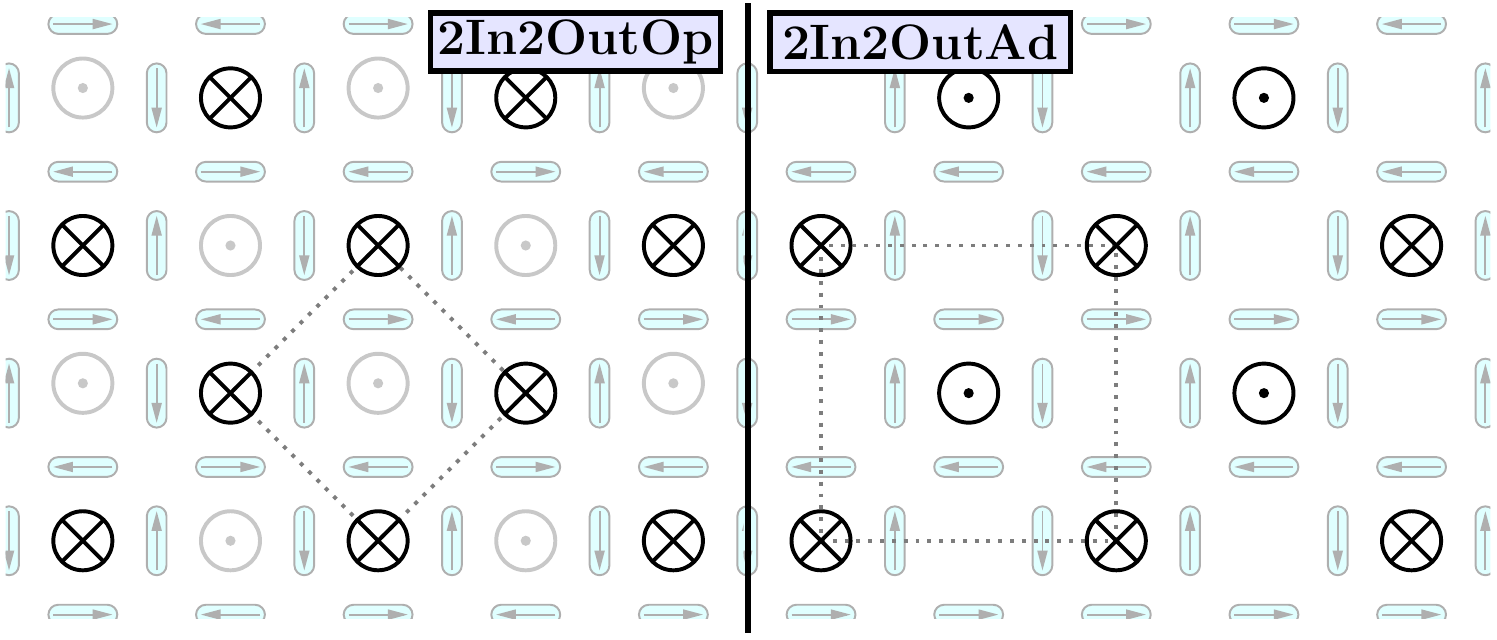}
 \caption{(Color online) Ground states in a square-lattice dipolar array. Circles and dotted lines represent flux closures and magnetic unit cells, respectively. Left: `2In2OutOp' configuration with $c(2\times 2)$ magnetic unit cell. The vortex in the center of the unit cell shows opposite chirality (brighter circles) with respect to those at the corners. Right: `2In2OutAd' configuration with $p(2\times 2)$ magnetic unit cell. Plaquettes at the edge centers of the unit cell do not exhibit flux closure.}
 \label{fig:vortex}
\end{figure}

The `2In2OutOp' configuration imposes a vortex structure described by a $c(2\times 2)$ magnetic unit cell (cf.\ the checkerboard structure on the left-hand side of Fig.~\ref{fig:vortex}). While for `2In2OutOp' a vortex chirality can be ascribed to each plaquette, this is not the case for the `2In2OutAd' configuration. Consequently, the latter displays a $p(2 \times 2)$ magnetic unit cell (on the right-hand side of Fig.~\ref{fig:vortex}).

By displacing vertically the rows and columns of the square lattice, the degeneracies of the above configurations can be tuned to $6$ for a critical $\delta z_{\mathrm{c}}$ (Ref.~\onlinecite{ThonigArXiv13}), that is, to the same degree of degeneracy as in water ice\cite{Pauling35}, in pyrochlore lattices\cite{Bramwell01,Matsuhira02}, and in kagom\'e lattices\cite{Huegli12} (In the honeycomb and square lattice this sixfold degeneracy is out of 8 and 16 possible vertices, respectively. However, one may consider both lattices and their magnetic ground states equivalent because both have the same residual entropy of $0.2\, k_{\mathrm{B}}$; cf.\ Ref.~\onlinecite{ThonigArXiv13}). Consequently, a transition from $c(2 \times 2)$ to $p(2 \times 2)$ ordering takes place at this critical displacement. For $\delta z < \delta z_{\mathrm{c}}$, `2in2outOp' nodes prevail, while for $\delta z > \delta z_{\mathrm{c}}$, `2in2outAd' vortices prevail. The critical value $\delta z_{\mathrm{c}}$ depends on the islands' shape: $\delta z = 0.27\, a$ for the realistic shapes used in this investigation (cf.\ Ref.~\onlinecite{ThonigArXiv13}), $0.418\, a$ in Ref.~\onlinecite{Moeller06}, and $0.444\, a$ in Ref.~\onlinecite{Mol10}.

\section{Discussion of results}
\label{sec:discussion}

In what follows, we distinguish two types of arrays: $\mathcal{S}_{1}$ has $\delta_{z} = 0$, a vanishing residual (zero-temperature) entropy and exhibits the `2In2OutOp' ground state. In contrast, $\mathcal{S}_{2}$ has the critical $\delta_{z} = 0.27\, a$, shows a residual, finite entropy of $S = 0.2\, k_{\mathrm{B}}$ (Ref.~\onlinecite{Morris09}) and its ground state is comprised of `2In2OutOp' and `2In2OutAd' vortices. The temperature is $T = \unit[300]{K}$ (room temperature).

\subsection{Randomly distributed vacancies}
\label{sec:vacancies}
For discussing the role of vacancies, we distribute randomly vacancies by removing nanoislands from dipolar arrays of type $\mathcal{S}_{1}$ with $N_{\mathrm{spin}} = 5000$. The vacancy concentration is $c \in [0, 0.5]$. Nodes at which one or three islands are left introduce charges $Q \in \{-3, -1, 1, 3 \}$; these nodes are not considered in the following, for we focus on nodes with charges of $\pm 4$. 

\begin{figure}
 \centering
 \includegraphics[width = 0.89\columnwidth]{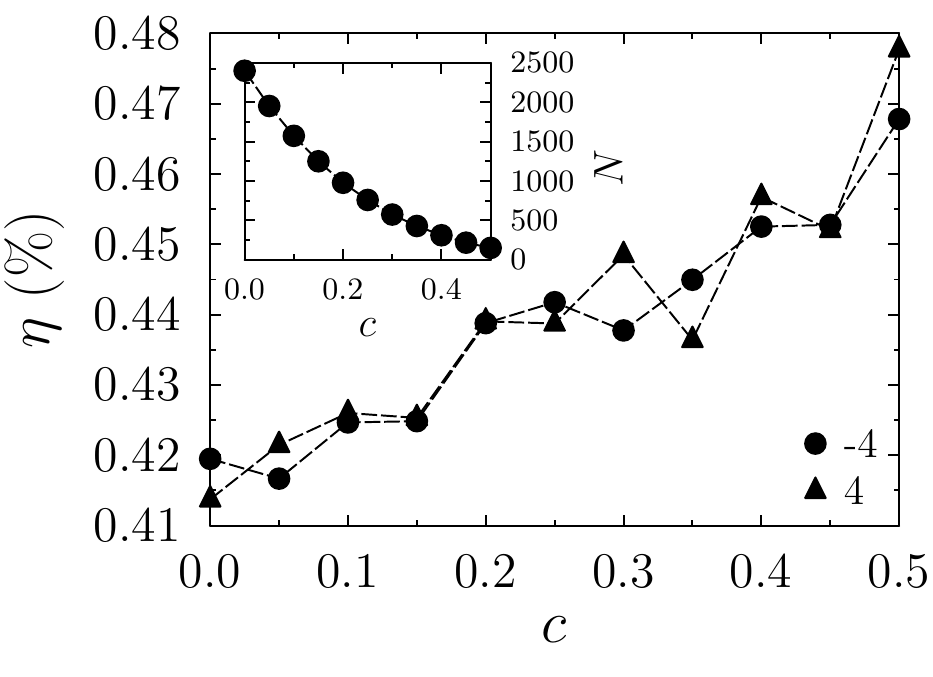}
 \caption{Effect of randomly distributed vacancies on the number of excitations with charge $-4$ (filled circles) and $+4$ (filled triangles). The fractions $\eta_{\pm 4}$, normalized to the number $N$ of nodes with four islands, are shown versus the vacancy concentration $c$. The inset depicts the reduction of $N$ with concentration $c$. Lines are guides to the eye.}
 \label{fig:vacancies}
\end{figure}

Removing a single island from a node reduces the energy of this defect by a factor of $2$ with respect to the unchanged node; the degree of degeneracy remains unaltered. It turns out that nodes with reduced exchange coupling among the islands result in a minute increase of the number of nodes with $Q = \pm 4$. More precisely, the fractions of charges $\eta_{\pm4}$ increase with the concentration $c$ by about $\unit[0.07]{\%}$ (Fig.~\ref{fig:vacancies}; the data shown are averages over $30$ ensembles and 100 Monte Carlo runs).

The minute increase of $\eta_{\pm 4}$ is attributed to the exponentially reduced number of contributing nodes, i.\,e., those with four islands (inset in Fig.~\ref{fig:vacancies}). The roughly linear behavior of $\eta_{\pm 4}(c)$ suggests that defective nodes and nodes next to the defects do not longer obey the ice rule; consequently, transitions between close energy levels appear. An analysis of the charge correlation function which probes the number of pairs of nodes with a distance $r$ (not shown here) yields no spatial correlation between high magnetic excitations and vacancies.

In a recent investigation, Silva et al. \cite{Silva13} report that low magnetic charges (e.\,g., two in/one out and one in/two out vortex states) pinned at vacancies obey strong Coulomb interactions.  A second energy contribution, introduced \textit{ad hoc}, describes the interaction of string excitations with the defect. The Coulomb-type and the \textit{ad hoc} contribution are in the order of $\unit[0.01-0.1]{meV}$ and, thus, too small to be resolved at room temperature (which corresponds to a thermal energy of about $\unit[30]{meV}$).

Thus, we conclude that a distribution of vacancies, either random or controlled, does neither produce a significant enhancement nor allows control of thermally excited charges $\pm 4$. Therefore, we now turn to defects with modified island properties.

\subsection{Defects with modified island properties}
\label{sec:defects}
The degree of frustration is a driving factor for enlarged ratios $\eta_{\pm 4}$ (e.\,g., shown in Ref.~\onlinecite{ThonigArXiv13}). This suggests to add defects, that is, nodes with modified properties with respect to the ideal dipolar array, at specified positions and with prescribed properties. These defect nodes comprise nanoislands with modified magnetization density, thickness or vertical displacement (Section~\ref{sec:theory} and Fig.~\ref{fig:defect}). If these  modified nodes would pin excitations with charges $\pm 4$, one could expect string excitations that connect a pair of defects. In the following, $N_{\mathrm{spin}} = 800$.

\begin{figure}
 \centering
 \includegraphics[width = 0.87\columnwidth]{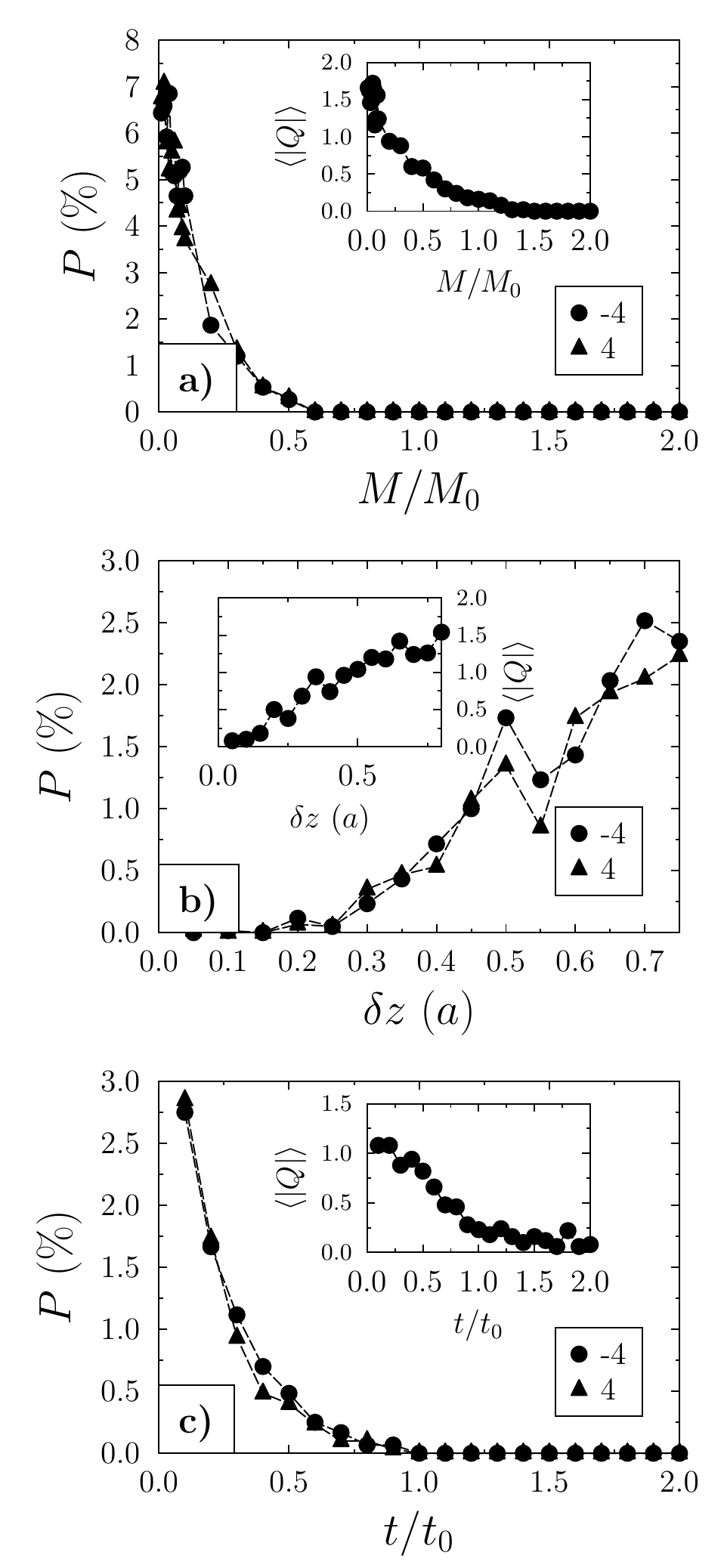}
 \caption{Pinning of excitations with charges $+4$ (filled circles) and $-4$ (filled triangles) at defects in a dipolar lattice of type $\mathcal{S}_{1}$ with (a)
 varied magnetization $M$, (b) vertical displacement $\delta z$, and (c) modified island's thickness $t$. 
 The probability $P$ is given versus $M$, $\delta z$, and $t$, with respect to the quantities of the ideal dipolar array $M_{0}$, $\delta z = 0$, and $t_{0} = \unit[3]{nm}$. The insets depict the average $\langle |Q| \rangle$ of the absolute charge at a defect. Temperature $T = \unit[300]{K}$. Lines are guides to the eye. }
 \label{fig:properties}
\end{figure}

First, we show that defects pin charges of $\pm 4$. The probability $P$ of finding a magnetic charge $\pm 4$ at a defect depends on the defect's properties: the magnetization $M$, the vertical displacement $\delta z$, and the islands' thickness $t$. For a lattice of type $\mathcal{S}_{1}$ it increases exponentially up to $\unit[3]{\%}$ (Fig.~\ref{fig:properties}b and c) or even up to $\unit[8]{\%}$ (Fig.~\ref{fig:properties}a). A decrease of $M$ and $c$ results in an increase of $P$. The opposite behavior is found for $\delta z$, a finding corroborating that a lattice with a globally increased $\delta z$ shows larger fractions $\eta_{\pm 4}$.  The data in Fig.~\ref{fig:properties} are averages over $100$ Monte Carlo sets and $60$ ensembles.

In view of the entire dipolar array, the probability of finding a $Q = \pm 4$ node at a defect is two orders of magnitude larger than that for finding such a node in the rest of the system ($\eta_{\mathrm{tot}} \approx \unit[0.03]{\%}$). This finding supports that defects are efficient pinning centers and corroborates the \textit{ad hoc} interaction between defects and magnetic charges motivated in Ref.~\onlinecite{Silva13}. Furthermore, the mean absolute value $\langle |Q| \rangle$ for a defect yields that magnetic charges $\pm 2$ are most likely (insets in Fig.~\ref{fig:properties}), in accordance with experiment \cite{Morgan10}. Hence, the spin-ice rule does not apply even for minor deviations from the ideal lattice and at finite temperature.

The next step is to show that excitations with $Q = \pm 4$ are simultaneously present at a pair of defects.
To illuminate this issue, we choose two defects, D1 and D2, with magnetization $M  = 0.1\, M_{0}$ (cf.\ Fig.~\ref{fig:properties}a) positioned on a straight line along the rows or columns. The distance between the defects is $d$. A quantitative analysis is provided by  the charge-correlation function $S_{Q}(r, d)$ which is the probability of finding a charge $Q$ in a distance $r$ from the defect D1 for a given $d$.

\begin{figure}
 \centering
 \includegraphics[width = 1.0\columnwidth]{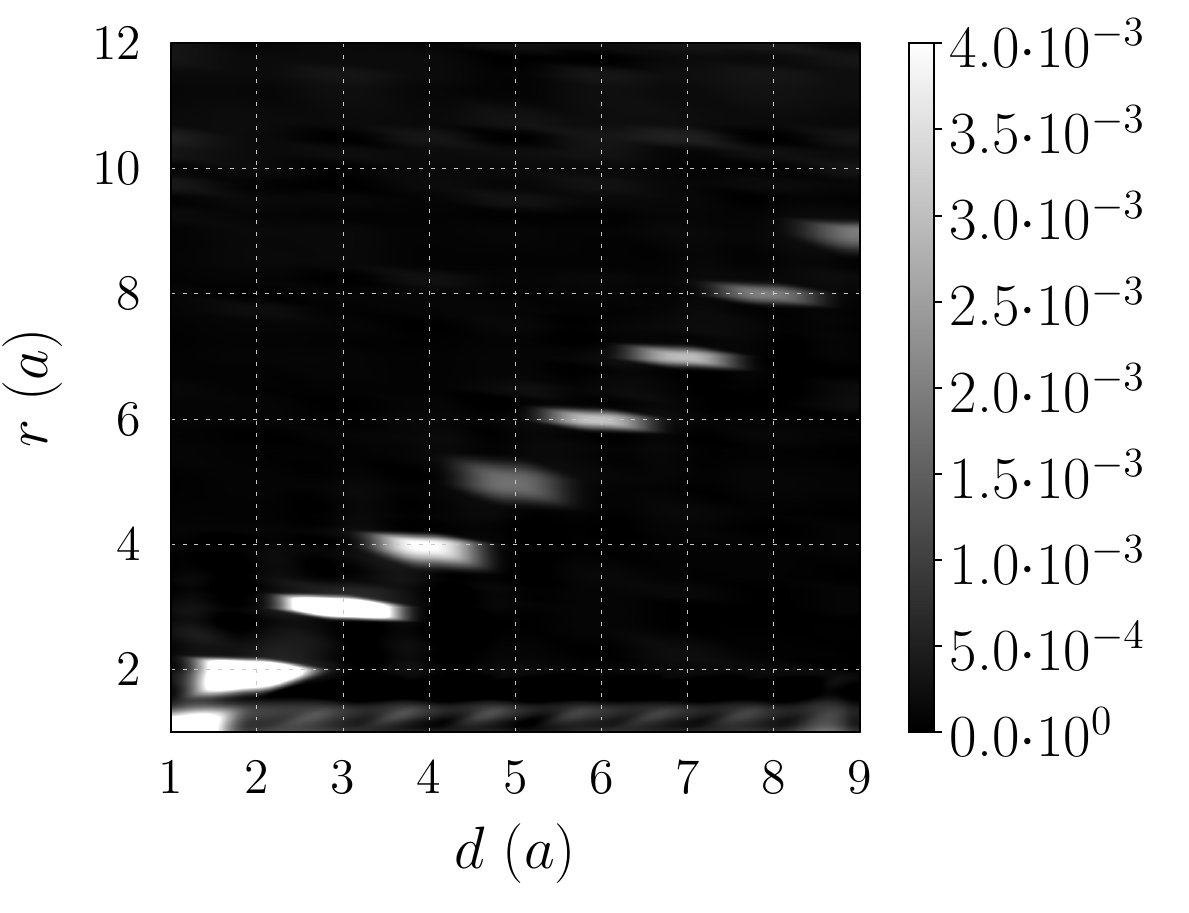}
 \caption{Charge correlation function $S_{\pm 4}(r, d)$ for charges $Q = - 4$ vs.\ distance $d$ between two defects and distance $r$ from the first defect, depicted as gray scale (right). $a$ is the lattice constant. Because $S_{+4} = S_{-4}$ on average, data for $Q = +4$ are not shown. A mesh is added to guide the eye. The finite size of the spots is due to interpolation of the computed data at the mesh points.}
 \label{fig:distancedefects}
\end{figure}

$S_{\pm 4}(r, d)$ shows two maxima (Fig.~\ref{fig:distancedefects}). The largest peak is at $r = 0$ (not shown), indicating the trivial `self-correlation' of a defect. More importantly, maxima appear at $r = d$, that is, charges $Q = \pm 4$ show up simultaneously at the two defects. Furthermore, the probability of finding thermal excitations with $Q = \pm 4$ and $Q = \pm 2$ (not shown here) at the defects is enhanced with respect to the ideal array. From the amplitude of this peak in $S$ we deduce a  correlation length of up to four nearest-neighbor distances, in agreement with Ref.~\onlinecite{Morgan10}. On top of the above correlation analysis we investigated the charge correlation function for oppositely charged defects. Again, we find maxima at $r = d$. Please recall that string excitations are links between oppositely charged nodes (Fig.~\ref{fig:string}).

The charge correlation function $S$ makes no statement about a ferromagnetic string connecting the two defects, as has been found in kagom\'e and square-lattice spin ices \cite{Mengotti11,Morgan10}. However, the maxima at $r = d$ strongly suggest such links. In the following we will discuss string excitations connecting two defects.

\begin{figure}
 \centering
 \includegraphics[width = \columnwidth]{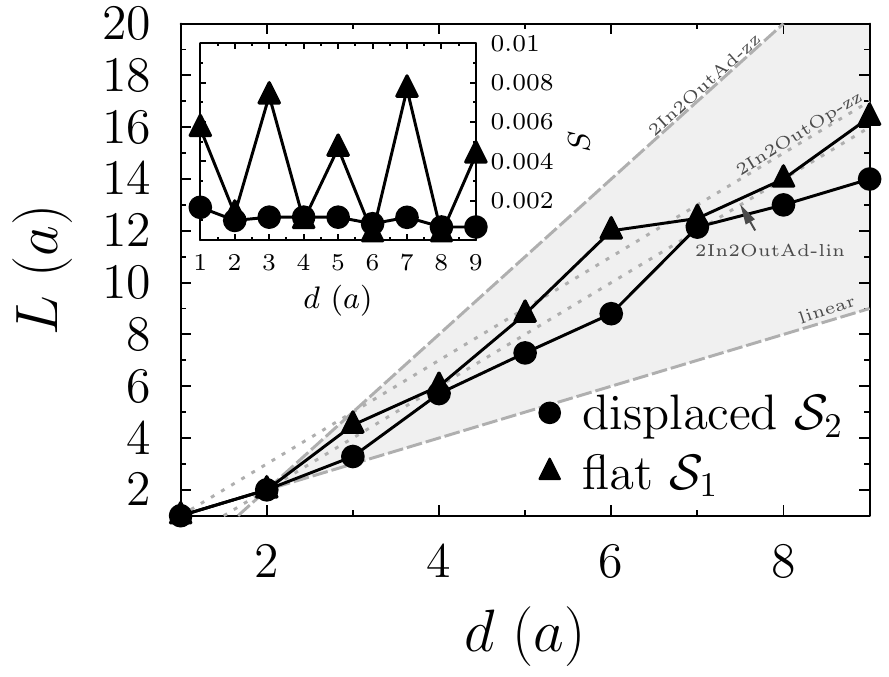}
 \caption{Average minimal length $L$ of string excitations versus distance $d$ between two defects ($a$ lattice constant). Data are given for `flat' ($\mathcal{S}_{1}$, filled triangles) and vertically displaced ($\mathcal{S}_{2}$, filled circles) dipolar arrays. The dotted lines are analytically derived minimal string lengths \textit{(i)} for a linear string (`linear'), \textit{(ii)} for a `zig-zag' path in the `2In2OutOp' and '2In2OutAd' states (`2In2OutOp-zz' and `2In2OutAd-zz'), and \textit{(iii)} for a straight line offset by $a$ with respect to the direct connection (`2In2OutAd-lin'). `Minimal' strings appear within the shaded area. The inset shows the probability of finding a string with the minimal length $L$.}
 \label{fig:lengthdiracstring}
\end{figure}

As suggested above, we find that two defects are connected by string excitations. The string length $L$ is approximately twice as large as the distance of the defects $d$ (Fig.~\ref{fig:lengthdiracstring}). This result is explained by the topology of the square array. In arrays of type $\mathcal{S}_{1}$ with zero residual entropy, the dominant `2In2OutOp' configuration induces a $c(2 \times 2)$ arrangement of vortices (Fig.~\ref{fig:vortex}a)\footnote{In the limit of $L \gg d$, the number of string configurations is given by the random walk result $3^{\nicefrac{L}{a}}$, where $a$ is the lattice constant. We use an $n$-ary tree scheme\cite{tree} to obtain the shortest ferromagnetic string by iteratively increasing the length $L$.}. Thus, the  shortest link between two defects is a `zig-zag' path (cf.\ Fig.~\ref{fig:string}). For defects at a distance $d$ which is an odd multiple of the lattice constant $a$, this `zig-zag' is compatible with the $c(2 \times 2)$ structure. For example, defects at a distance of $d = 3\, a$ are joined by a string of minimal length $L = 5\, a$, whereas those at a distance of  $d = 5\, a$ by $L = 9\, a$: $L = 2 d - a$. In contrast, defects at distances that are even multiples require a breaking of the flux closure near the defects, which in turn increases the string energy. As a consequence, the probability of finding shortest strings is large for defects at an `odd' distance, but small for defects at an `even' distance (inset in Fig.~\ref{fig:lengthdiracstring}).

The above topological restriction does not hold for arrays of type $\mathcal{S}_{2}$ with nonzero residual entropy because
the `2In2OutOp' and `2In2OutAd' configurations are equally likely. As a result, such samples show both $c(2\times 2)$ and $p(2\times 2)$ arrangements of the vortices (Fig.~\ref{fig:vortex}). A `minimal' string is then a straight line that is offset by one column next to the direct connection of the defects; its length is $L = d + 2\,a$.  Note that a direct connection, with $L = d$, is unlikely because of the $c(2\times 2)$ arrangement of the flux closure.

As suggested by M\'ol et al.\ (Ref.~\onlinecite{Mol09}), the energetics include a nonzero string tension, which is described by a potential $V(d) = Q/ d + b L + c$. $b$ is the string tension, and $c$ is associated with the monopole-pair creation. Consequently, a `minimal' string minimizes the energy. For deducing the string tension for a given sample at a finite temperature, the system should be thermally stable; this suggests to use type $\mathcal{S}_{1}$ rather than the thermally active type $\mathcal{S}_{2}$ (cf.\ Refs.~\onlinecite{Mol09}, \onlinecite{Mol10}, and \onlinecite{Mol12} for a discussion). We have compared the average energy with and without defects for both types $\mathcal{S}_{1}$ and $\mathcal{S}_{2}$. In accordance with Castelnovo et al. \cite{Castelnovo08} and M\'ol et al. \cite{Mol09}, a maximum shows up at about $d = 9\, a$. A shift of the energy maximum which is related to the Coulomb contribution from other strings was not observed, in contrast to a prediction by Silva et al. \cite{Silva13}. This is explained by the marginal probability to find other $\pm 4$ charges in the $\mathcal{S}_{1}$ dipolar array. Moreover, the string tension `softens' with temperature, thus, confirming a statement given in Ref.~\onlinecite{Mol10}. The thermal energy of about $\unit[30]{meV}$ (room temperature is chosen for this Paper) is too large to resolve the Coulomb-type contribution $Q/d$ (Ref.~\onlinecite{Mol09}) which is in the order of $\unit[0.1]{meV}$. 

Eventually, we address briefly the modification of the defects. Since all modifications change the dipolar interaction energies (with respect to the ideal array), one can obtain prescribed energies by all modifications discussed in this Paper. Analytical calculations for which we assume islands without lateral extension, yield that the magnetization fulfills $\nicefrac{M}{M_{0}} = \nicefrac{t}{2 t_{0}}$, where $M_{0}$ and $t_{0}$ are the magnetization and the thickness of an unperturbed island, respectively. Furthermore, the vertical displacement $\delta z$ and the magnetization are linked: $\delta z = \beta \sqrt{\left(\nicefrac{M_{0}}{M}\right)^{\nicefrac{4}{5}} - 1}$, with $\beta = 1$. Numerical calculations for realistic island shapes give $\beta = 0.37$ and a minute deviation from the above linearity in $M$ versus $t$. 

\section{Concluding remarks}
\label{sec:conclusion}
In this theoretical investigations we have shown that thermal string excitations can be pinned at modified nanomagnets in a dipolar array. The different types of defects may be produced by microstructuring techniques. It turned out that a decrease of the islands' magnetization density is most efficient in creating nodes with large charges. On top of this, a pair of defects is connected by a ferromagnetic path of islands, that is, by a string excitation.

The length of the pinned strings is closely related to magnetic ground state configuration and to the position of the defects. This finding suggests experimental and theoretical investigations of topology and formation of magnetic domains.

\acknowledgments
We thank Stephan Rei{\ss}aus for fruitful discussions.

\bibliographystyle{apsrev}

\end{document}